\documentclass[12pt]{article}
\usepackage{geometry}
\usepackage{graphicx}
\usepackage{ragged2e,microtype}
\usepackage[utf8]{inputenc}
    \usepackage{times}
\usepackage{enumitem,amssymb}
\usepackage{pifont}
\usepackage{mathptmx}  
\usepackage[colorlinks=true,citecolor=blue, urlcolor=blue, linkcolor=blue]{hyperref}
\usepackage{upgreek}
\usepackage{xspace}
\usepackage{natbib}

\newlist{thematic}{itemize}{8}
\setlist[thematic]{label=$\square$}

\def\mum{\ensuremath{\upmu\mathrm{m}}\xspace}

\makeatletter
\let\jnl@style=\rm
\def\ref@jnl#1{{\jnl@style#1}}

\def\aj{\ref@jnl{AJ}}                   
\def\actaa{\ref@jnl{Acta Astron.}}      
\def\araa{\ref@jnl{ARA\&A}}             
\def\apj{\ref@jnl{ApJ}}                 
\def\apjl{\ref@jnl{ApJ}}                
\def\apjs{\ref@jnl{ApJS}}               
\def\ao{\ref@jnl{Appl.~Opt.}}           
\def\apss{\ref@jnl{Ap\&SS}}             
\def\aap{\ref@jnl{A\&A}}                
\def\aapr{\ref@jnl{A\&A~Rev.}}          
\def\aaps{\ref@jnl{A\&AS}}              
\def\azh{\ref@jnl{AZh}}                 
\def\baas{\ref@jnl{BAAS}}               
\def\bac{\ref@jnl{Bull. astr. Inst. Czechosl.}}
\def\caa{\ref@jnl{Chinese Astron. Astrophys.}}
\def\cjaa{\ref@jnl{Chinese J. Astron. Astrophys.}}
\def\icarus{\ref@jnl{Icarus}}           
\def\jcap{\ref@jnl{J. Cosmology Astropart. Phys.}}
\def\jrasc{\ref@jnl{JRASC}}             
\def\memras{\ref@jnl{MmRAS}}            
\def\mnras{\ref@jnl{MNRAS}}             
\def\na{\ref@jnl{New A}}                
\def\nar{\ref@jnl{New A Rev.}}          
\def\pra{\ref@jnl{Phys.~Rev.~A}}        
\def\prb{\ref@jnl{Phys.~Rev.~B}}        
\def\prc{\ref@jnl{Phys.~Rev.~C}}        
\def\prd{\ref@jnl{Phys.~Rev.~D}}        
\def\pre{\ref@jnl{Phys.~Rev.~E}}        
\def\prl{\ref@jnl{Phys.~Rev.~Lett.}}    
\def\pasa{\ref@jnl{PASA}}               
\def\pasp{\ref@jnl{PASP}}               
\def\pasj{\ref@jnl{PASJ}}               
\def\qjras{\ref@jnl{QJRAS}}             
\def\rmxaa{\ref@jnl{Rev. Mexicana Astron. Astrofis.}}%
\def\rnaas{\ref@jnl{RNAAS}}             
\def\skytel{\ref@jnl{S\&T}}             
\def\solphys{\ref@jnl{Sol.~Phys.}}      
\def\sovast{\ref@jnl{Soviet~Ast.}}      
\def\ssr{\ref@jnl{Space~Sci.~Rev.}}     
\def\zap{\ref@jnl{ZAp}}                 
\def\nat{\ref@jnl{Nature}}              
\def\natas{\ref@jnl{Nat.~Ast.}}         
\def\iaucirc{\ref@jnl{IAU~Circ.}}       
\def\aplett{\ref@jnl{Astrophys.~Lett.}} 
\def\apspr{\ref@jnl{Astrophys.~Space~Phys.~Res.}}
\def\bain{\ref@jnl{Bull.~Astron.~Inst.~Netherlands}} 
\def\fcp{\ref@jnl{Fund.~Cosmic~Phys.}}  
\def\gca{\ref@jnl{Geochim.~Cosmochim.~Acta}}   
\def\grl{\ref@jnl{Geophys.~Res.~Lett.}} 
\def\jcp{\ref@jnl{J.~Chem.~Phys.}}      
\def\jgr{\ref@jnl{J.~Geophys.~Res.}}    
\def\jqsrt{\ref@jnl{J.~Quant.~Spec.~Radiat.~Transf.}}
\def\memsai{\ref@jnl{Mem.~Soc.~Astron.~Italiana}}
\def\nphysa{\ref@jnl{Nucl.~Phys.~A}}   
\def\physrep{\ref@jnl{Phys.~Rep.}}   
\def\physscr{\ref@jnl{Phys.~Scr}}   
\def\planss{\ref@jnl{Planet.~Space~Sci.}}   
\def\procspie{\ref@jnl{Proc.~SPIE}}   

\def\rprphys{\ref@jnl{Rep.~Prog.~Phys\@}}   
\def\ptp{\ref@jnl{Prog.~Th.~Phys.}}   
\def\natas{\ref@jnl{NatAs}}           
\def\amjm{\ref@jnl{AmJM}}             

\makeatother

\begin{document}
\raggedright
\huge
Building a Roadmap for Hubble Science into the 2030s\linebreak

Revealing Atmospheric Structure and Evolution in Substellar Worlds Using HST \linebreak
\normalsize


Name:	Allison M. McCarthy	
 \linebreak					
Institution:  Trinity College Dublin
 \linebreak 
E-mail: \href{mailto:mccara45@tcd.ie}{mccara45@tcd.ie}
 \\
 
\textbf{Co-authors:} (names and institutions) \\

Merle A. Schrader (Trinity College Dublin)
\\
Johanna M. Vos (Trinity College Dublin)
\\
Sven Kiefer (University of Texas at Austin)
\\
Cian O'Toole (Trinity College Dublin)
\\
Michael K. Plummer (United States Air Force Academy)
\\
Michael Poon (University of Toronto)
\\
Daniella Bardalez Gagliuffi (Amherst College)
\\
Samuel Beiler (Trinity College Dublin)
\\
John E. Gizis (University of Delaware)
\\
Melodie M. Kao (Lowell Observatory)
\\
Gabriel-Dominique Marleau (Universit\"at Duisburg--Essen)
\\
Elisabeth C. Matthews (Max Planck Institute for Astronomy)
\\
Philip S. Muirhead (Boston University)
\\
Evert Nasedkin (Trinity College Dublin)
\\
Natalia Oliveros-Gomez (Johns Hopkins University)
\\
J. Sebastian Pineda (University of Colorado Boulder LASP)
\\
Kimberly Ward-Duong (Smith College)
\\

\textbf{Abstract  (optional):}

Substellar objects occupy a unique place in our universe, bridging the gap between the smallest stars and the largest planets, and serving as powerful laboratories for understanding extrasolar atmospheric physics without the contaminating glare of a host star. Previous studies into the atmospheric structure of these objects have revealed clouds, disequilibrium chemistry, thermal inversions, and auroral processes which each contribute to wavelength-dependent brightness variations. HST remains uniquely positioned to address key open questions in the field, such as resolving the vertical atmospheric structure, long term evolution of the atmosphere, detection of UV aurora in the upper atmosphere, primarily in conjunction with other facilities that probe wavelength regimes that cannot be reached with instruments on HST.  We advocate for three large scale initiatives and argue that the study of the atmospheres of substellar worlds directly prepares the community for atmospheric characterization with the Habitable Worlds Observatory.

\pagebreak
\justifying
\section*{Introduction}
\vspace{-0.5em}

Substellar worlds -- L, T, and Y type dwarfs -- bridge the gap between the smallest stars and the largest planets, placing them in a unique position in our universe. These objects often overlap with exoplanets in terms of mass, age, and temperature \citep{Faherty2016}, but unlike directly imaged exoplanets they are not contaminated by the glare of their bright host stars, allowing high precision atmospheric characterization across a wide sample. Rotational variability monitoring has shown that substellar atmospheres are spatially inhomogeneous. As different atmospheric regions rotate in and out of view, they produce wavelength-dependent brightness variations that probe distinct atmospheric altitudes and pressure levels, allowing the reconstruction of three-dimensional atmospheric structure \citep[e.g.][]{Artigau2009, Radigan20122M2139, Morley2014, Tremblin2016}. Four physical mechanisms are thought to drive this variability, each operating across distinct atmospheric layers and timescales, and each observable over distinct wavelength ranges:

\textbf{Clouds and Hazes:} Cloud decks form at pressures and temperatures where silicate, salt, sulfide, and iron species condense. With decreasing effective temperature, these cloud structures become increasingly spatially inhomogeneous as condensates sediment (rain) below the photosphere  \citep[e.g.][]{AckermanMarley2001,Marley2010}. Clouds obscure the intrinsic thermal emission from the brown dwarf, leading to rotational brightness variations as the object rotates \citep[e.g.][]{Apai2013,Metchev2015}. Whether the observed variability arises primarily from vertically coherent cloud structures \citep{Tan&Showman2019}, localized cloud gaps, or temperature perturbations that indirectly modify cloud formation \citep{Nasedkin2025}, remains unclear.

\textbf{Chemistry:} Substellar atmospheres are not in chemical equilibrium. Vertical mixing transports gas between atmospheric layers faster than chemical reactions can re-establish equilibrium abundances, resulting in enhanced and depleted molecular species relative to equilibrium. The clearest evidence of this is the presence of CO and CO$_2$ features in T and Y dwarfs, where carbon is expected to be primarily locked up in CH$_4$ \citep{Lodders2002,Geballe2009,Sorahana2012,Beiler2024a}. The turbulent processes responsible for this mixing are also expected to produce spectral variability, particularly at 3--5~\mum \citep{McCarthy2025}. However, despite the clear observational evidence for vertical mixing, the physical drivers and variability of this process are poorly understood. 

\textbf{Upper Atmosphere Heating and Thermal Inversions:} Additional energy sources, including dynamical processes and auroras can heat the upper atmosphere faster than it can radiatively equilibrate, producing thermal inversions in which temperature increases with altitude \citep[e.g.][]{Morley2014, Pineda2017}. Such inversions have only recently been identified in several extrasolar atmospheres \citep[e.g.][]{McCarthy2025,Nasedkin2025,Chen2025}, but their feedback on cloud formation and chemistry is poorly understood.

\textbf{Aurorae:} Several brown dwarfs are known to host auroral processes, identified through radio emission generated by the electron cyclotron maser instability \citep{RouteandWolszczan2012,Kao2016,Kao2018}. Auroral energy deposition into the upper atmosphere may contribute to the thermal inversions described above, but disentangling auroral heating from other sources of upper atmospheric heating has proven difficult \citep[e.g.][]{McCarthy2025,Nasedkin2025}. Jupiter's aurora is observed both in the infrared and the ultraviolet (UV), and extrasolar aurorae are expected to behave similarly, though likely with different energy partitioning due to the differences in atmospheric composition and density. However, since substellar objects emit most strongly in the infrared, infrared auroral signatures are expected to be difficult to disentangle from atmospheric variability caused by clouds or temperature perturbations. UV aurorae provide a cleaner diagnostic because brown dwarfs emit negligible intrinsic ultraviolet thermal radiation \citep{Saur2021}.

\vspace{-0.5em}

\subsection*{Beyond Atmospheric Characterization}

Rotational variability monitoring can also address questions beyond atmospheric characterization. Rotation periods measured through rotational variability monitoring of wide-separation, planetary-mass companions can constrain their spin-orbit alignment \citep{Bryan+2020_Obliquity}, distinguish between planet-like formation pathways that produce aligned systems and stellar or dynamical formation mechanisms that generate large obliquities \citep{Poon+2025}. Rotation rate distributions as a function of age and mass is also a valuable dataset to test evolutionary models and angular momentum evolution \citep{Bryan2018}. Both obliquities and substellar evolution models inform our understanding of how planetary systems form and evolve.

\begin{figure}[]
    \centering
    \includegraphics[width=0.88\linewidth]{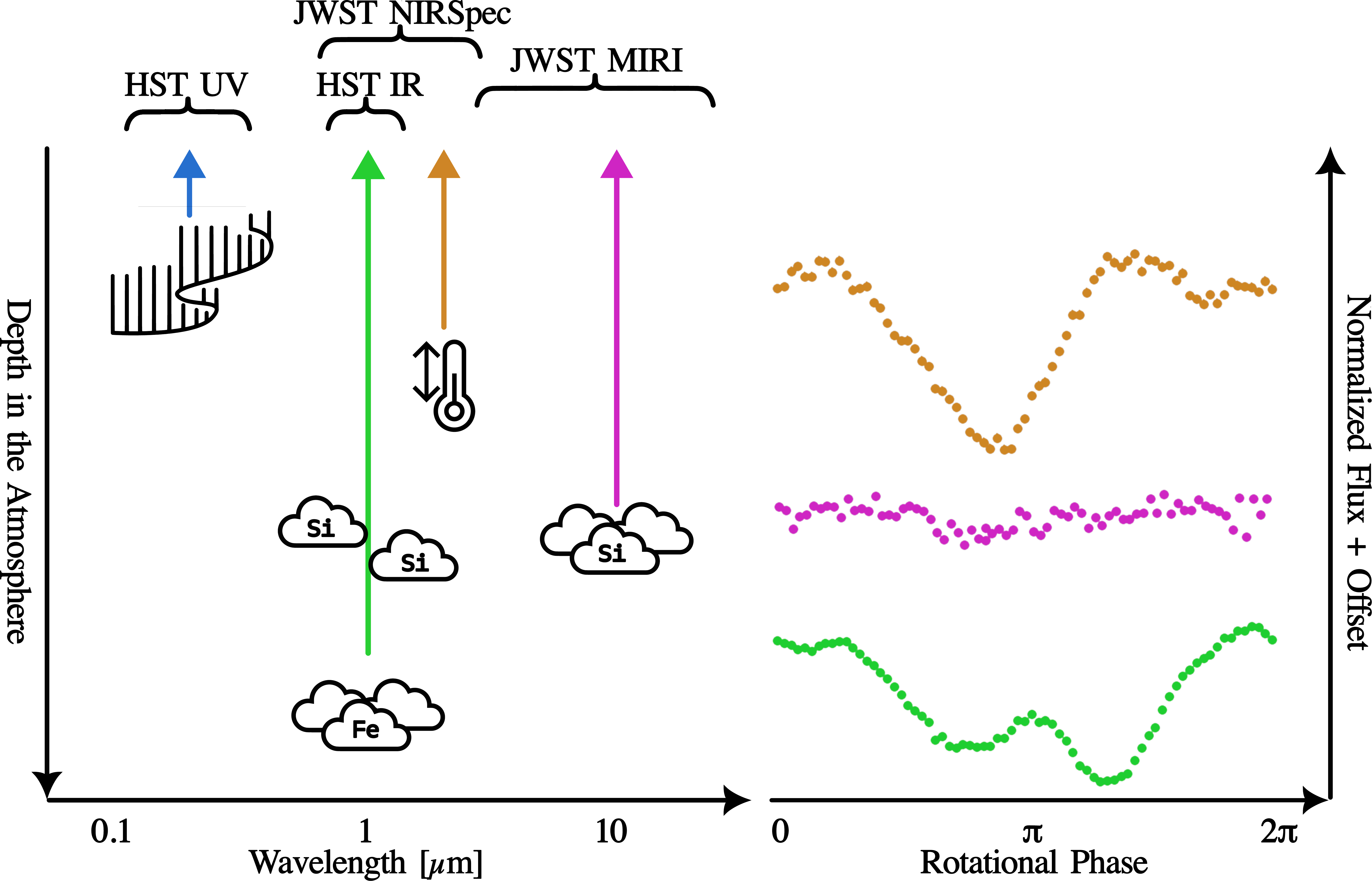}
    \caption{\textbf{Left:} Schematic illustration of the primary physical drivers of variability in substellar atmospheres (from left to right: aurora, iron clouds obscured by patchy silicate clouds, thermal inversion, and silicate clouds), their relative atmospheric depths and the wavelengths at which their signatures are most clearly observed. The instruments capable of probing these wavelengths and their drivers are shown at the top of the illustration. \textbf{Right:} Rotational light curves from the young, bright T2.5 brown dwarf SIMP J013656.5+093347 from GO 3548 (PI: Vos). Near-infrared variability traces deep cloud layers and higher-atmosphere temperature inversions (green and orange respectively, JWST NIRSpec PRISM), while the mid-infrared feature probes intermediate atmospheric cloud layers (pink, JWST MIRI LRS). JWST requires sequential observations with different instruments and observing modes, the NIR and MIR observations sample different rotations. Ultraviolet auroral emission is expected to originate at the highest atmospheric altitudes but has not yet been detected outside our Solar System.}
   \label{fig:contrFunc}
   \vspace{-0.5em}
\end{figure}

\section{HST's Role in Understanding Substellar Atmospheres}

HST transformed our understanding of substellar atmospheres through high-precision, time-resolved near-infrared spectroscopy, demonstrating that variability amplitudes and phases are strongly wavelength-dependent and that atmospheric structures extend through multiple pressure levels \citep[e.g.][]{Buenzli2012,Yang2016}. Yet, several objects exhibit variability inconsistent with any single mechanism \citep[e.g.][]{Tremblin2015,Biller2018}, and rotation-to-rotation light curve evolution suggests atmospheric circulation changes on short timescales \citep{Apai2017}. Critical questions remain unanswered regarding the vertical structure and evolution of substellar atmospheres, and HST -- particularly in combination with JWST and the VLA -- remains uniquely positioned to address them.

\textbf{Joint HST/WFC3 + JWST/MIRI: Resolving Vertical Atmospheric Structure} 

Clouds, chemistry, and upper atmosphere heating all manifest across a wide range of wavelengths and pressure levels, and disentangling their contributions requires spanning from the near-IR to the mid-IR in a single atmospheric snapshot. Although HST/WFC3 and JWST/NIRSpec both cover 0.8--1.7~\mum (Figure \ref{fig:contrFunc}), neither instrument alone can fully resolve the vertical structure of the atmosphere. HST/WFC3's near-IR grism spectroscopy can probe deep into the atmosphere and water absorption bands, constraining cloud and haze particle sizes through the Rayleigh slope, and providing clues into the distribution and inhomogeneity of cloud layers, while JWST/MIRI access the $\sim 10$\mum silicate feature \citep[e.g.][]{Biller2024}, constraining cloud mass, thickness, and composition \citep[e.g.][]{Welbanks2024, Changeat2025, Kiefer2026}. 

The critical operational requirement is strict simultaneity. Obtaining near- and mid-IR variability with JWST alone requires sequential observations with different instruments, meaning resulting light curves sample different rotational phases. Because brown dwarf atmospheres evolve from rotation to rotation \citep{Apai2017}, sequential observations cannot be interpreted as a single vertically resolved atmospheric snapshot.

\textbf{Long-Baseline Monitoring with HST/WFC3: Tracking Atmospheric Evolution}

Snapshot observations reveal atmospheric structure at a single epoch, and short term monitoring captures evolution across at most several consecutive rotations. Understanding how clouds, chemistry, and thermal inversions evolve over months to years instead requires repeated observations over long baselines. Repeated one- or two-orbit HST visits to the same target over extended baselines will directly measure how the atmosphere evolves across many rotations. Existing long-baseline constraints are largely derived from low-resolution ground-based observations lacking time-resolved photometry using different facilities, filters and observing modes, leading to a lack of overlap between observation techniques, observed wavelengths, and instrumental sensitivities. Using HST WFC3 to answer the question of long term evolution resolves these discrepancies.

The key instrumental requirement for this set of observations is flux calibration stability between visits. Repeat-visit monitoring programs must maintain sufficiently stable flux calibration to distinguish genuine atmospheric evolution from instrumental systematics. Without this stability, long-baseline programs cannot reliably separate real variability from visit-to-visit calibration differences. Current reduced-gyro operations of HST additionally motivate efficient and flexible monitoring strategies. Rather than relying entirely on long uninterrupted HST sequences as we have in the past, repeated short visits provide a scientifically valuable and operationally practical method for long-baseline observations. This positions HST and JWST as compliments -- JWST delivers detailed continuous variability, while HST tracks how the climate shapes those same atmospheres over substantially longer baselines.

\textbf{HST/STIS: The Only Window to UV Auroral Emission}

Ultraviolet auroral detection on substellar worlds is only possible with HST. Brown dwarfs emit negligible intrinsic ultraviolet thermal radiation \citep{Saur2021}, making UV emission a clean tracer of auroral energy deposition and magnetospheric dynamics, uncontaminated by the cloud and thermal variability that complicates infrared auroral searches. Previous searches may not have reached required sensitivities for potentially variable auroral emission, while differences in the observed auroral traces to those on Jupiter further complicate interpretation \citep{Saur2021, Pineda2024}.

Understanding how auroral energy is deposited and redistributed through extrasolar atmospheres requires multi-wavelength observations. UV observations directly trace auroral heating, while simultaneous radio observations confirm magnetospheric activity. 
HST/STIS, particularly with its \texttt{TIME-TAG} mode,  
is the only observatory capable of accessing the 1100--1700~\AA\ wavelength range that is sensitive enough for the detection and monitoring of these aurorae, making STIS irreplaceable for this science case. The key operational requirement is scheduling flexibility that allows for simultaneous HST UV and radio observations with facilities such as the VLA.

\section{Preparation for the Habitable Worlds Observatory}

The Hubble Space Telescope can critically support the development of the Habitable Worlds Observatory (HWO) by serving as a testbed for its advanced scientific capabilities. The time-resolved spectroscopic variability techniques developed over the last 30 years such as light curve decomposition, multi-wavelength phase analysis, and calibration strategies for repeat-visit programs will continue to be honed in the next decade. These techniques are the same methods that HWO will use to characterize variability in directly imaged exoplanet atmospheres. Continued HST programs provide an unparalleled opportunity to polish these techniques before HWO begins operations.

HWO is also uniquely positioned to observe UV aurora on exoplanets \citep{HWO_UV_Aurora}, making UV auroral characterization a priority science case for HST. HST UV campaigns which study radio-loud substellar objects now will allow the community to develop and refine detection strategies, sensitivity requirements, and analysis techniques that HWO will rely on. 

Finally, the practical insights gained from HST monitoring programs can directly inform HWO instrument design. Atmospheric variability monitoring of faint substellar objects requires precise flux calibration stability between observations, sufficient spectral resolution to disentangle molecular features and atmospheric layers, and precise wavelength coverage to optimize the combination of variability mechanisms conserved. The lessons learned from years of HST observations will be critical in optimizing the instruments aboard the HWO.

\section{Proposed Large Scale Observing Programs}

Although some of the science questions laid out above are better suited to targeted single object observations, others are well suited to large-scale community programs that would generate legacy datasets of broad value.
\vspace{-0.5em}
\subsection{Long Term Monitoring Initiative}

Substellar atmospheres are known to evolve from rotation to rotation \citep{Apai2017}, but global circulation models also suggest long term evolution over the course of months to years \citep{Tan2025}. Currently, there is a lack of observational evidence against which to test these predictions, and existing repeated observations of the same target often use different instruments at different facilities with different observing strategies. A dedicated HST program of repeated short visits to a carefully selected sample of substellar objects — spanning effective temperature, age, and mass — using a consistent instrumental setup would directly measure how atmospheres evolve across many rotations. An observing program of this magnitude would allow for consistent observations of isolated objects with the same instrument and observing mode would provide observational evidence to back global circulation model predictions.

\subsection{The Search for Extrasolar Aurora}

Around 17--25\% of M dwarfs, 10--13\% of L dwarfs, and 23--29\% of T/Y dwarfs display quiescent radio emission that likely traces radiation belts \citep{Kao2023,Climent2023} and is linked to aurorae \citep{Pineda2017,Kao2019,Kao2023,Kao2024AuroraOccurrence}. In the solar system, aurorae on outer planets manifest in the radio, infrared, and ultraviolet \citep[e.g.][]{GrodentJupiterAuroraUV,NicholsJupiterAuroraIR} and extrasolar auroras are expected to manifest in the same wavelengths \citep{Pineda2017,Saur2021}. However, despite the high occurrence rate of magnetospheric emissions, UV extrasolar aurorae have not yet been confirmed. A large scale HST UV campaign targeting known radio loud substellar objects spanning the full L/T/Y sequence would be a significant step forward by providing the first systematic and sufficiently sensitive search for extrasolar UV aurora, directly addressing one of the key open questions in substellar atmospheric science, while simultaneously building the foundation for the detection and analysis of aurora using the HWO.

\subsection{Benchmark Brown Dwarfs in Old Open Clusters}

Young, low-gravity substellar objects receive significant attention because they exhibit higher occurrence rates and amplitudes of atmospheric variability \citep{Vos2020}. As a result, the sample of old, age-benchmarked counterparts remains small. This data gap is problematic because as brown dwarfs age, they cool and contract, increasing their surface gravity and triggering the formation of complex condensate clouds. Although theoretical models predict how these older, cooler atmospheres behave, empirical testing is impossible without age-benchmarked targets. A large-scale observing program designed to discover, classify, and study the atmospheric structure of  substellar objects in old, open clusters would effectively fill this critical observational gap.

\pagebreak

{\footnotesize{
\bibliographystyle{yahapj} 
\bibliography{bibfile}}}

@ARTICLE{Metchev2015,
       author = {{Metchev}, Stanimir A. and {Heinze}, Aren and {Apai}, D{\'a}niel and
         {Flateau}, Davin and {Radigan}, Jacqueline and {Burgasser}, Adam and
         {Marley}, Mark S. and {Artigau}, {\'E}tienne and {Plavchan}, Peter and
         {Goldman}, Bertrand},
        title = "{Weather on Other Worlds. II. Survey Results: Spots are Ubiquitous on L and T Dwarfs}",
      journal = {\apj},
         year = "2015",
        month = "Feb",
       volume = {799},
       number = {2},
          eid = {154},
        pages = {154},
          doi = {10.1088/0004-637X/799/2/154},
archivePrefix = {arXiv},
       eprint = {1411.3051},
 primaryClass = {astro-ph.SR},
       adsurl = {https://ui.adsabs.harvard.edu/abs/2015ApJ...799..154M}
}

@ARTICLE{Tan&Showman2019,
       author = {{Tan}, Xianyu and {Showman}, Adam P.},
        title = "{Atmospheric Variability Driven by Radiative Cloud Feedback in Brown Dwarfs and Directly Imaged Extrasolar Giant Planets}",
      journal = {\apj},
         year = 2019,
        month = apr,
       volume = {874},
       number = {2},
          eid = {111},
        pages = {111},
          doi = {10.3847/1538-4357/ab0c07},
archivePrefix = {arXiv},
       eprint = {1809.06467},
 primaryClass = {astro-ph.SR},
       adsurl = {https://ui.adsabs.harvard.edu/abs/2019ApJ...874..111T}
}

@article{AckermanMarley2001,
  author={Ackerman, A. S. and Marley, M. S.},
  title={Precipitating condensation clouds in substellar atmospheres},
  year={2001},
  journal={Astrophys. J.},
  volume={556},
  pages={872--884},
  doi={10.1086/321540},
}

@ARTICLE{GrodentJupiterAuroraUV,
       author = {{Grodent}, Denis and {Bonfond}, B. and {Yao}, Z. and {G{\'e}rard}, J.-C. and {Radioti}, A. and {Dumont}, M. and {Palmaerts}, B. and {Adriani}, A. and {Badman}, S.~V. and {Bunce}, E.~J. and {Clarke}, J.~T. and {Connerney}, J.~E.~P. and {Gladstone}, G.~R. and {Greathouse}, T. and {Kimura}, T. and {Kurth}, W.~S. and {Mauk}, B.~H. and {McComas}, D.~J. and {Nichols}, J.~D. and {Orton}, G.~S. and {Roth}, L. and {Saur}, J. and {Valek}, P.},
        title = "{Jupiter's Aurora Observed With HST During Juno Orbits 3 to 7}",
      journal = {Journal of Geophysical Research (Space Physics)},
         year = 2018,
        month = may,
       volume = {123},
       number = {5},
        pages = {3299-3319},
          doi = {10.1002/2017JA025046},
       adsurl = {https://ui.adsabs.harvard.edu/abs/2018JGRA..123.3299G}
}

@ARTICLE{NicholsJupiterAuroraIR,
       author = {{Nichols}, J.~D. and {King}, O.~R.~T. and {Clarke}, J.~T. and {de Pater}, I. and {Fletcher}, L.~N. and {Melin}, H. and {Moore}, L. and {Tao}, C. and {Yeoman}, T.~K.},
        title = "{Dynamic infrared aurora on Jupiter}",
      journal = {Nature Communications},
         year = 2025,
        month = may,
       volume = {16},
       number = {1},
          eid = {3907},
        pages = {3907},
          doi = {10.1038/s41467-025-58984-z},
       adsurl = {https://ui.adsabs.harvard.edu/abs/2025NatCo..16.3907N}
}

@ARTICLE{Tan2025,
       author = {{Tan}, Xianyu and {Zhang}, Xi and {Marley}, Mark S. and {Zhou}, Yifan and {Lew}, Ben W.~P. and {Miles}, Brittany E. and {Batalha}, Natasha E. and {Biller}, Beth A. and {Chauvin}, Ga{\"e}l and {Hinkley}, Sasha and {Hoch}, Kielan K.~W. and {Manjavacas}, Elena and {Metchev}, Stanimir and {Petrus}, Simon and {Rickman}, Emily and {Skemer}, Andrew and {Su{\'a}rez}, Genaro and {Sutlieff}, Ben J. and {Vos}, Johanna M. and {Whiteford}, Niall},
        title = "{Large-amplitude variability driven by giant dust storms on a planetary-mass companion}",
      journal = {Science Advances},
         year = 2025,
        month = nov,
       volume = {11},
        pages = {22.3324},
          doi = {10.1126/sciadv.adv3324},
archivePrefix = {arXiv},
       eprint = {2511.23163},
 primaryClass = {astro-ph.EP},
       adsurl = {https://ui.adsabs.harvard.edu/abs/2025SciA...11v3324T}
}

@ARTICLE{Changeat2025,
       author = {{Changeat}, Q. and {Bardet}, D. and {Chubb}, K. and {Dyrek}, A. and {Edwards}, B. and {Ohno}, K. and {Venot}, O.},
        title = "{Cloud and haze parameterization in atmospheric retrievals: Insights from Titan's Cassini data and JWST observations of hot Jupiters}",
      journal = {\aap},
         year = 2025,
        month = jul,
       volume = {699},
          eid = {A219},
        pages = {A219},
          doi = {10.1051/0004-6361/202453186},
archivePrefix = {arXiv},
       eprint = {2505.18715},
 primaryClass = {astro-ph.EP},
       adsurl = {https://ui.adsabs.harvard.edu/abs/2025A&A...699A.219C}
}

@ARTICLE{Kiefer2026,
       author = {{Kiefer}, Sven and {Morley}, Caroline V. and {Rowland}, Melanie J.},
        title = "{Connecting JWST Silicate Cloud Observations to Exoplanet Cloud Microphysics with Nimbus}",
      journal = {\apj},
         year = 2026,
        month = apr,
       volume = {1001},
       number = {1},
          eid = {98},
        pages = {98},
          doi = {10.3847/1538-4357/ae5101},
archivePrefix = {arXiv},
       eprint = {2603.13167},
 primaryClass = {astro-ph.EP},
       adsurl = {https://ui.adsabs.harvard.edu/abs/2026ApJ..1001...98K}
}

@ARTICLE{RouteandWolszczan2012,
       author = {{Route}, M. and {Wolszczan}, A.},
        title = "{The Arecibo Detection of the Coolest Radio-flaring Brown Dwarf}",
      journal = {\apjl},
         year = 2012,
        month = mar,
       volume = {747},
       number = {2},
          eid = {L22},
        pages = {L22},
          doi = {10.1088/2041-8205/747/2/L22},
archivePrefix = {arXiv},
       eprint = {1202.1287},
 primaryClass = {astro-ph.SR},
       adsurl = {https://ui.adsabs.harvard.edu/abs/2012ApJ...747L..22R}
}

@ARTICLE{Bryan2018,
       author = {{Bryan}, Marta L. and {Benneke}, Bj{\"o}rn and {Knutson}, Heather A. and {Batygin}, Konstantin and {Bowler}, Brendan P.},
        title = "{Constraints on the spin evolution of young planetary-mass companions}",
      journal = {Nature Astronomy},
         year = 2018,
        month = dec,
       volume = {2},
        pages = {138-144},
          doi = {10.1038/s41550-017-0325-8},
archivePrefix = {arXiv},
       eprint = {1712.00457},
 primaryClass = {astro-ph.EP},
       adsurl = {https://ui.adsabs.harvard.edu/abs/2018NatAs...2..138B}
}

@ARTICLE{Kao2023,
       author = {{Kao}, Melodie M. and {Mioduszewski}, Amy J. and {Villadsen}, Jackie and {Shkolnik}, Evgenya L.},
        title = "{Resolved imaging confirms a radiation belt around an ultracool dwarf}",
      journal = {\nat},
         year = 2023,
        month = jul,
       volume = {619},
       number = {7969},
        pages = {272-275},
          doi = {10.1038/s41586-023-06138-w},
archivePrefix = {arXiv},
       eprint = {2302.12841},
 primaryClass = {astro-ph.EP},
       adsurl = {https://ui.adsabs.harvard.edu/abs/2023Natur.619..272K}
}

@ARTICLE{Climent2023,
       author = {{Climent}, J.~B. and {Guirado}, J.~C. and {P{\'e}rez-Torres}, M. and {Marcaide}, J.~M. and {Pe{\~n}a-Mo{\~n}ino}, L.},
        title = "{Evidence for a radiation belt around a brown dwarf}",
      journal = {Science},
         year = 2023,
        month = sep,
       volume = {381},
       number = {6662},
        pages = {1120-1124},
          doi = {10.1126/science.adg6635},
archivePrefix = {arXiv},
       eprint = {2303.06453},
 primaryClass = {astro-ph.SR},
       adsurl = {https://ui.adsabs.harvard.edu/abs/2023Sci...381.1120C}
}

@ARTICLE{Kao2019,
       author = {{Kao}, Melodie M. and {Hallinan}, Gregg and {Pineda}, J. Sebastian},
        title = "{Constraints on magnetospheric radio emission from Y dwarfs}",
      journal = {\mnras},
         year = 2019,
        month = aug,
       volume = {487},
       number = {2},
        pages = {1994-2004},
          doi = {10.1093/mnras/stz1372},
       adsurl = {https://ui.adsabs.harvard.edu/abs/2019MNRAS.487.1994K}
}

@ARTICLE{Kao2024AuroraOccurrence,
       author = {{Kao}, Melodie M. and {Shkolnik}, Evgenya L.},
        title = "{The occurrence rate of quiescent radio emission for ultracool dwarfs using a generalized semi-analytical Bayesian framework}",
      journal = {\mnras},
         year = 2024,
        month = jan,
       volume = {527},
       number = {3},
        pages = {6835-6866},
          doi = {10.1093/mnras/stad2272},
archivePrefix = {arXiv},
       eprint = {2306.16460},
 primaryClass = {astro-ph.SR},
       adsurl = {https://ui.adsabs.harvard.edu/abs/2024MNRAS.527.6835K}
}

@ARTICLE{Nasedkin2025,
       author = {{Nasedkin}, E. and {Schrader}, M. and {Vos}, J.~M. and {Biller}, B. and {Burningham}, B. and {Cowan}, N.~B. and {Faherty}, J.~K. and {Gonzales}, E. and {Lam}, M.~B. and {McCarthy}, A.~M. and {Muirhead}, P.~S. and {O'Toole}, C. and {Plummer}, M.~K. and {Su{\'a}rez}, G. and {Tan}, X. and {Visscher}, C. and {Whiteford}, N. and {Zhou}, Y.},
        title = "{The JWST weather report: Retrieving temperature variations, auroral heating, and static cloud coverage on SIMP-0136}",
      journal = {\aap},
         year = 2025,
        month = oct,
       volume = {702},
          eid = {A1},
        pages = {A1},
          doi = {10.1051/0004-6361/202555370},
archivePrefix = {arXiv},
       eprint = {2507.07772},
 primaryClass = {astro-ph.EP},
       adsurl = {https://ui.adsabs.harvard.edu/abs/2025A&A...702A...1N}
}

@ARTICLE{Apai2017,
   author = {{Apai}, D. and {Karalidi}, T. and {Marley}, M.~S. and {Yang}, H. and 
	{Flateau}, D. and {Metchev}, S. and {Cowan}, N.~B. and {Buenzli}, E. and 
	{Burgasser}, A.~J. and {Radigan}, J. and {Artigau}, E. and {Lowrance}, P.
	},
    title = "{Zones, spots, and planetary-scale waves beating in brown dwarf atmospheres}",
  journal = {Science},
     year = 2017,
    month = aug,
   volume = 357,
    pages = {683-687},
      doi = {10.1126/science.aam9848},
   adsurl = {https://ui.adsabs.harvard.edu/abs/2017Sci...357..683A}
}

@ARTICLE{Vos2020,
       author = {{Vos}, Johanna M. and {Biller}, Beth A. and {Allers}, Katelyn N. and {Faherty}, Jacqueline K. and {Liu}, Michael C. and {Metchev}, Stanimir and {Eriksson}, Simon and {Manjavacas}, Elena and {Dupuy}, Trent J. and {Janson}, Markus and {Radigan-Hoffman}, Jacqueline and {Crossfield}, Ian and {Bonnefoy}, Micka{\"e}l and {Best}, William M.~J. and {Homeier}, Derek and {Schlieder}, Joshua E. and {Brandner}, Wolfgang and {Henning}, Thomas and {Bonavita}, Mariangela and {Buenzli}, Esther},
        title = "{Spitzer Variability Properties of Low-gravity L Dwarfs}",
      journal = {\aj},
         year = 2020,
        month = jul,
       volume = {160},
       number = {1},
          eid = {38},
        pages = {38},
          doi = {10.3847/1538-3881/ab9642},
archivePrefix = {arXiv},
       eprint = {2005.12854},
 primaryClass = {astro-ph.SR},
       adsurl = {https://ui.adsabs.harvard.edu/abs/2020AJ....160...38V}
}

@ARTICLE{Tremblin2015,
       author = {{Tremblin}, P. and {Amundsen}, D.~S. and {Mourier}, P. and {Baraffe}, I. and {Chabrier}, G. and {Drummond}, B. and {Homeier}, D. and {Venot}, O.},
        title = "{Fingering Convection and Cloudless Models for Cool Brown Dwarf Atmospheres}",
      journal = {\apjl},
         year = 2015,
        month = may,
       volume = {804},
       number = {1},
          eid = {L17},
        pages = {L17},
          doi = {10.1088/2041-8205/804/1/L17},
archivePrefix = {arXiv},
       eprint = {1504.03334},
 primaryClass = {astro-ph.SR},
       adsurl = {https://ui.adsabs.harvard.edu/abs/2015ApJ...804L..17T}
}

@ARTICLE{Artigau2009,
       author = {{Artigau}, {\'E}tienne and {Bouchard}, Sandie and {Doyon}, Ren{\'e} and {Lafreni{\`e}re}, David},
        title = "{Photometric Variability of the T2.5 Brown Dwarf SIMP J013656.5+093347: Evidence for Evolving Weather Patterns}",
      journal = {\apj},
         year = 2009,
        month = aug,
       volume = {701},
       number = {2},
        pages = {1534-1539},
          doi = {10.1088/0004-637X/701/2/1534},
archivePrefix = {arXiv},
       eprint = {0906.3514},
 primaryClass = {astro-ph.SR},
       adsurl = {https://ui.adsabs.harvard.edu/abs/2009ApJ...701.1534A}
}

@ARTICLE{Yang2016,
       author = {{Yang}, Hao and {Apai}, D{\'a}niel and {Marley}, Mark S. and {Karalidi}, Theodora and {Flateau}, Davin and {Showman}, Adam P. and {Metchev}, Stanimir and {Buenzli}, Esther and {Radigan}, Jacqueline and {Artigau}, {\'E}tienne and {Lowrance}, Patrick J. and {Burgasser}, Adam J.},
        title = "{Extrasolar Storms: Pressure-dependent Changes in Light-curve Phase in Brown Dwarfs from Simultaneous HST and Spitzer Observations}",
      journal = {\apj},
         year = 2016,
        month = jul,
       volume = {826},
       number = {1},
          eid = {8},
        pages = {8},
          doi = {10.3847/0004-637X/826/1/8},
archivePrefix = {arXiv},
       eprint = {1605.02708},
 primaryClass = {astro-ph.EP},
       adsurl = {https://ui.adsabs.harvard.edu/abs/2016ApJ...826....8Y}
}

@ARTICLE{Kao2016,
       author = {{Kao}, Melodie M. and {Hallinan}, Gregg and {Pineda}, J. Sebastian and {Escala}, Ivanna and {Burgasser}, Adam and {Bourke}, Stephen and {Stevenson}, David},
        title = "{Auroral Radio Emission from Late L and T Dwarfs: A New Constraint on Dynamo Theory in the Substellar Regime}",
      journal = {\apj},
         year = 2016,
        month = feb,
       volume = {818},
       number = {1},
          eid = {24},
        pages = {24},
          doi = {10.3847/0004-637X/818/1/24},
archivePrefix = {arXiv},
       eprint = {1511.03661},
 primaryClass = {astro-ph.SR},
       adsurl = {https://ui.adsabs.harvard.edu/abs/2016ApJ...818...24K}
}

@ARTICLE{Apai2013,
       author = {{Apai}, D{\'a}niel and {Radigan}, Jacqueline and {Buenzli}, Esther and {Burrows}, Adam and {Reid}, Iain Neill and {Jayawardhana}, Ray},
        title = "{HST Spectral Mapping of L/T Transition Brown Dwarfs Reveals Cloud Thickness Variations}",
      journal = {\apj},
         year = 2013,
        month = may,
       volume = {768},
       number = {2},
          eid = {121},
        pages = {121},
          doi = {10.1088/0004-637X/768/2/121},
archivePrefix = {arXiv},
       eprint = {1303.4151},
 primaryClass = {astro-ph.EP},
       adsurl = {https://ui.adsabs.harvard.edu/abs/2013ApJ...768..121A}
}

@ARTICLE{Tremblin2016,
       author = {{Tremblin}, P. and {Amundsen}, D.~S. and {Chabrier}, G. and {Baraffe}, I. and {Drummond}, B. and {Hinkley}, S. and {Mourier}, P. and {Venot}, O.},
        title = "{Cloudless Atmospheres for L/T Dwarfs and Extrasolar Giant Planets}",
      journal = {\apjl},
         year = 2016,
        month = feb,
       volume = {817},
       number = {2},
          eid = {L19},
        pages = {L19},
          doi = {10.3847/2041-8205/817/2/L19},
archivePrefix = {arXiv},
       eprint = {1601.03652},
 primaryClass = {astro-ph.EP},
       adsurl = {https://ui.adsabs.harvard.edu/abs/2016ApJ...817L..19T}
}

@ARTICLE{Biller2018,
       author = {{Biller}, Beth A. and {Vos}, Johanna and {Buenzli}, Esther and {Allers}, Katelyn and {Bonnefoy}, Micka{\"e}l and {Charnay}, Benjamin and {B{\'e}zard}, Bruno and {Allard}, France and {Homeier}, Derek and {Bonavita}, Mariangela and {Brandner}, Wolfgang and {Crossfield}, Ian and {Dupuy}, Trent and {Henning}, Thomas and {Kopytova}, Taisiya and {Liu}, Michael C. and {Manjavacas}, Elena and {Schlieder}, Joshua},
        title = "{Simultaneous Multiwavelength Variability Characterization of the Free-floating Planetary-mass Object PSO J318.5-22}",
      journal = {\aj},
         year = 2018,
        month = feb,
       volume = {155},
       number = {2},
          eid = {95},
        pages = {95},
          doi = {10.3847/1538-3881/aaa5a6},
archivePrefix = {arXiv},
       eprint = {1712.03746},
 primaryClass = {astro-ph.EP},
       adsurl = {https://ui.adsabs.harvard.edu/abs/2018AJ....155...95B}
}

@ARTICLE{Buenzli2012,
       author = {{Buenzli}, Esther and {Apai}, D{\'a}niel and {Morley}, Caroline V. and {Flateau}, Davin and {Showman}, Adam P. and {Burrows}, Adam and {Marley}, Mark S. and {Lewis}, Nikole K. and {Reid}, I. Neill},
        title = "{Vertical Atmospheric Structure in a Variable Brown Dwarf: Pressure-dependent Phase Shifts in Simultaneous Hubble Space Telescope-Spitzer Light Curves}",
      journal = {\apjl},
         year = 2012,
        month = dec,
       volume = {760},
       number = {2},
          eid = {L31},
        pages = {L31},
          doi = {10.1088/2041-8205/760/2/L31},
archivePrefix = {arXiv},
       eprint = {1210.6654},
 primaryClass = {astro-ph.SR},
       adsurl = {https://ui.adsabs.harvard.edu/abs/2012ApJ...760L..31B}
}

@ARTICLE{Morley2014,
       author = {{Morley}, Caroline V. and {Marley}, Mark S. and {Fortney}, Jonathan J. and {Lupu}, Roxana},
        title = "{Spectral Variability from the Patchy Atmospheres of T and Y Dwarfs}",
      journal = {\apjl},
         year = 2014,
        month = jul,
       volume = {789},
       number = {1},
          eid = {L14},
        pages = {L14},
          doi = {10.1088/2041-8205/789/1/L14},
archivePrefix = {arXiv},
       eprint = {1406.0863},
 primaryClass = {astro-ph.SR},
       adsurl = {https://ui.adsabs.harvard.edu/abs/2014ApJ...789L..14M}
}

@ARTICLE{Faherty2016,
       author = {{Faherty}, Jacqueline K. and {Riedel}, Adric R. and {Cruz}, Kelle L. and {Gagne}, Jonathan and {Filippazzo}, Joseph C. and {Lambrides}, Erini and {Fica}, Haley and {Weinberger}, Alycia and {Thorstensen}, John R. and {Tinney}, C.~G. and {Baldassare}, Vivienne and {Lemonier}, Emily and {Rice}, Emily L.},
        title = "{Population Properties of Brown Dwarf Analogs to Exoplanets}",
      journal = {\apjs},
         year = 2016,
        month = jul,
       volume = {225},
       number = {1},
          eid = {10},
        pages = {10},
          doi = {10.3847/0067-0049/225/1/10},
archivePrefix = {arXiv},
       eprint = {1605.07927},
 primaryClass = {astro-ph.SR},
       adsurl = {https://ui.adsabs.harvard.edu/abs/2016ApJS..225...10F}
}

@ARTICLE{Biller2024,
       author = {{Biller}, Beth A. and {Vos}, Johanna M. and {Zhou}, Yifan and {McCarthy}, Allison M. and {Tan}, Xianyu and {Crossfield}, Ian J.~M. and {Whiteford}, Niall and {Suarez}, Genaro and {Faherty}, Jacqueline and {Manjavacas}, Elena and {Chen}, Xueqing and {Liu}, Pengyu and {Sutlieff}, Ben J. and {Limbach}, Mary Anne and {Molliere}, Paul and {Dupuy}, Trent J. and {Oliveros-Gomez}, Natalia and {Muirhead}, Philip S. and {Henning}, Thomas and {Mace}, Gregory and {Crouzet}, Nicolas and {Karalidi}, Theodora and {Morley}, Caroline V. and {Tremblin}, Pascal and {Kataria}, Tiffany},
        title = "{The JWST weather report from the nearest brown dwarfs I: multiperiod JWST NIRSpec + MIRI monitoring of the benchmark binary brown dwarf WISE 1049AB}",
      journal = {\mnras},
         year = 2024,
        month = aug,
       volume = {532},
       number = {2},
        pages = {2207-2233},
          doi = {10.1093/mnras/stae1602},
archivePrefix = {arXiv},
       eprint = {2407.09194},
 primaryClass = {astro-ph.SR},
       adsurl = {https://ui.adsabs.harvard.edu/abs/2024MNRAS.532.2207B}
}

@ARTICLE{Kao2018,
       author = {{Kao}, Melodie M. and {Hallinan}, Gregg and {Pineda}, J. Sebastian and {Stevenson}, David and {Burgasser}, Adam},
        title = "{The Strongest Magnetic Fields on the Coolest Brown Dwarfs}",
      journal = {\apjs},
         year = 2018,
        month = aug,
       volume = {237},
       number = {2},
          eid = {25},
        pages = {25},
          doi = {10.3847/1538-4365/aac2d5},
archivePrefix = {arXiv},
       eprint = {1808.02485},
 primaryClass = {astro-ph.SR},
       adsurl = {https://ui.adsabs.harvard.edu/abs/2018ApJS..237...25K}
}

@ARTICLE{Beiler2024a,
       author = {{Beiler}, Samuel A. and {Mukherjee}, Sagnick and {Cushing}, Michael C. and {Kirkpatrick}, J. Davy and {Schneider}, Adam C. and {Kothari}, Harshil and {Marley}, Mark S. and {Visscher}, Channon},
        title = "{A Tale of Two Molecules: The Underprediction of CO$_{2}$ and Overprediction of PH$_{3}$ in Late T and Y Dwarf Atmospheric Models}",
      journal = {\apj},
         year = 2024,
        month = sep,
       volume = {973},
       number = {1},
          eid = {60},
        pages = {60},
          doi = {10.3847/1538-4357/ad6759},
archivePrefix = {arXiv},
       eprint = {2407.15950},
 primaryClass = {astro-ph.EP},
       adsurl = {https://ui.adsabs.harvard.edu/abs/2024ApJ...973...60B}
}

@INPROCEEDINGS{Welbanks2024,
       author = {{Welbanks}, Luis and {Line}, Michael and {Carter}, Aarynn and {May}, Erin and {Powell}, Diana and {JWST Transiting Exoplanet Community ERS Team}},
        title = "{The first full broadband transmission spectrum of an exoplanet. The chemistry of WASP-39b revealed from 0.5 to 12 {\ensuremath{\mu}}m with JWST}",
    booktitle = {AAS/Division for Extreme Solar Systems Abstracts},
         year = 2024,
       series = {AAS/Division for Extreme Solar Systems Abstracts},
       volume = {56},
        month = apr,
          eid = {201.02},
        pages = {201.02},
       adsurl = {https://ui.adsabs.harvard.edu/abs/2024ESS.....520102W}
}

@ARTICLE{Pineda2017,
       author = {{Pineda}, J. Sebastian and {Hallinan}, Gregg and {Kao}, Melodie M.},
        title = "{A Panchromatic View of Brown Dwarf Aurorae}",
      journal = {\apj},
         year = 2017,
        month = sep,
       volume = {846},
       number = {1},
          eid = {75},
        pages = {75},
          doi = {10.3847/1538-4357/aa8596},
archivePrefix = {arXiv},
       eprint = {1708.02942},
 primaryClass = {astro-ph.SR},
       adsurl = {https://ui.adsabs.harvard.edu/abs/2017ApJ...846...75P}
}

@ARTICLE{Marley2010,
       author = {{Marley}, Mark S. and {Saumon}, Didier and {Goldblatt}, Colin},
        title = "{A Patchy Cloud Model for the L to T Dwarf Transition}",
      journal = {\apjl},
         year = 2010,
        month = nov,
       volume = {723},
       number = {1},
        pages = {L117-L121},
          doi = {10.1088/2041-8205/723/1/L117},
archivePrefix = {arXiv},
       eprint = {1009.6217},
 primaryClass = {astro-ph.SR},
       adsurl = {https://ui.adsabs.harvard.edu/abs/2010ApJ...723L.117M}
}

\end{document}